\documentclass[a4paper,11pt]{article}
\pdfoutput=1
\usepackage{pos}
\usepackage[firstpage]{draftwatermark}
\SetWatermarkFontSize{12pt}
\SetWatermarkColor{black}
\SetWatermarkAngle{0}
\SetWatermarkHorCenter{17.7cm}
\SetWatermarkVerCenter{5.8cm}
\SetWatermarkText{MS-TP-25-5, JLAB-THY-25-4211}
\usepackage{macros}
\usepackage{booktabs}
\usepackage{dsfont}
\usepackage{sidecap}
\usepackage{graphicx}
\newcommand\scalemath[2]{\scalebox{#1}{\mbox{\ensuremath{\displaystyle #2}}}}

\title{$\mathrm{O}(a)$ improvement of the flavour singlet scalar density\\ in a setup with Wilson fermions}
\ShortTitle{$\mathrm{O}(a)$ improvement of the flavour singlet scalar density}

\author*[a,b]{Pia Leonie Jones Petrak}
\author[b]{Jochen Heitger}
\author[b,c]{Fabian Joswig}
\affiliation[a]{Theory Center, Jefferson Lab, Newport News, Virginia 23606, USA}

\affiliation[b]{
	Institut für Theoretische Physik, Universität M\"unster,
	Wilhelm-Klemm-Straße 9, 48149  M\"unster, Germany}
\affiliation[c]{School of Physics and Astronomy,
	University of Edinburgh, Edinburgh EH9 3JZ, United Kingdom}

\emailAdd{pialjp@jlab.org}
\emailAdd{heitger@uni-muenster.de}

\abstract{We report on our Ward identity determination of the $\mathrm{O}(a)$ improvement coefficient for the flavour singlet scalar density, namely $g_\mathrm{S}$, from three-flavour lattice QCD with Wilson-clover fermions and the tree-level Symanzik improved
	gauge action. We employ five couplings, $g_0^2 \in [1.5,1.77]$, that cover the range used in large-volume CLS simulations. While $g_\mathrm{S}$ itself is for instance relevant for the $\mathrm{O}(a)$ improvement of meson and baryon sigma terms, a relation to $b_\mathrm{g}$, the $\mathrm{O}(a)$ improvement parameter of the gauge coupling, can also be established, allowing for its non-perturbative extraction as well. With Wilson fermions, $b_\mathrm{g}$ is in principle required for full $\mathrm{O}(a)$ improvement at non-vanishing sea quark masses. We outline our procedure for extracting $b_\mathrm{g}$.}

\FullConference{The 41st International Symposium on Lattice Field Theory (LATTICE2024)\\
 28 July - 3 August 2024\\
Liverpool, UK\\}

\begin{document}
\maketitle

\section{Introduction}

The $\mathrm{O}(a)$ improvement parameter of the flavour singlet scalar density, termed $\gs$, is for instance relevant for the full $\mathrm{O}(a)$ improvement of meson and baryon sigma terms, $\sigma_q=m_q \langle H|\bar{q}{q}|H\rangle$, in a setup with Wilson fermions, see \cite{Bhattacharya:2005rb}. The flavour singlet scalar density $\bar{q}q$ of flavour $q$ enters in the matrix element that is multiplied by the quark mass $m_q$. $|H\rangle$ refers to the ground state of a hadron $H$.  So far $\gs$ is only known from perturbation theory, see eq.~(\ref{eq:gs-pertubative}) below; as it is expected to be small it has been neglected in all sigma term calculations, as reported in \cite{FlavourLatticeAveragingGroupFLAG:2021npn,FlavourLatticeAveragingGroupFLAG:2024oxs}. In \cite{Agadjanov:2023efe} its impact is estimated from perturbation theory to be of the order of $-0.05\,\mathrm{MeV}$ on their estimate of the pion-nucleon sigma term of $\sigma_{\pi}^\mathrm{N}=(43.6\pm 3.8)\,\mathrm{MeV}$. Thus in this case perturbation theory indeed indicates that $\gs$ can be neglected since its effect is smaller than the uncertainty's. 

However, for a more reliable estimation, a non-perturbative study of $\gs$ is crucial and becoming more relevant as the precision of sigma term computations increases. The aim of our work is to perform a non-perturbative determination in the range of the CLS couplings, $g_0^2 \in [1.5,1.77]$.

Furthermore, the $\gs$ estimates may be used to obtain $\bg$ via the relation \cite{Bhattacharya:2005rb}
\begin{align}
	\bg = 2g_0^2 \gs \,.
	\label{bg-gs-relation}
\end{align}
In a setup with Wilson fermions the $\mathrm{O}(a)$ improvement parameter of the gauge coupling, $\bg$, is in principle required for full $\mathrm{O}(a)$ improvement at non-vanishing sea quark masses as $\tilde g_0^2 = \gosq \cdot ( 1 + ab_g(\gosq)  \Tr \Mq/\Nf )$ \cite{Bhattacharya:2005rb}, where $\Mq$ denotes the sea quark mass matrix. For the action and coupling range employed here, $\bg$ is only known from one-loop perturbation theory \cite{Luscher:1996sc,Sint:1997jx} as
\begin{align}
\label{eq:bg-pertubative}
b_\mathrm{g} = 0.012000(2) \times \Nf\, g_0^2 + \mathrm{O}\big(g_0^4\big)\,.
\end{align}
For a lower range of couplings, $g_0^2 \in [0.4,1.5]$, there has been a recent non-perturbative determination \cite{DallaBrida:2023fpl} relevant for ALPHA's decoupling strategy to determine the strong coupling \cite{DallaBrida:2022eua}. Our determination will be relevant for calculations using the CLS ensembles (see \cite{RQCD:2022xux}) such as sigma term determinations or other computations in the range of the CLS couplings with Wilson fermions.

By inserting eq.~(\ref{eq:bg-pertubative}) for $\bg$ in eq.~(\ref{bg-gs-relation}) we retrieve a lowest-order perturbative prediction for $\gs$,
\begin{align}
\label{eq:gs-pertubative}
\gs = 0.012000(2) \times \Nf/2 + \mathrm{O}\big(g_0^2\big)\,.
\end{align}
Note that this work's non-perturbative $\gs$ results cannot be used to extract $\bg$ via eq.~(\ref{bg-gs-relation}) and hence, cannot be directly compared to the above perturbative expression: an additional term needs to be included in the $\gs$ Ward identity, used here, for the relation from eq.~(\ref{bg-gs-relation}) to hold, see sect.~(\ref{sect:conclusion-and-outlook}). There, we also detail the next steps necessary to obtain the adequate $\gs$ estimates as to extract $\bg$ via eq.~(\ref{bg-gs-relation}).



\section{Chiral Ward Identities}

The well-known PCAC relation is derived from the general continuum axial Ward identity 
\begin{align}
\scalemath{0.9}{\int_\mathcal{R} \mathrm{d}^4x\hspace{0.1cm}\epsilon^a(x)\left[\partial_\mu\left<A_\mu^a(x)\mathcal{O}\right>-2m\left<P^a(x)\mathcal{O}\right>\right]
=-\left<\delta_\mathrm{A}\mathcal{O}\right>}
\label{general_axial}
\end{align}
by choosing the operator $\mathcal{O}$ to be an exterior single operator $\mathcal{O}_\mathrm{ext}$, i.e. located outside the region $R$ such that its axial variation is zero, $\left<\delta_\mathrm{A}\mathcal{O}\right>=\left<\delta_\mathrm{A}\mathcal{O}_\mathrm{ext}\right>=0$. Also, we always impose $\epsilon^a(x)=\epsilon=\mathrm{const.}$ for the smallness parameter so that the Ward identity reflects global chiral symmetry. $A_\mu^a$, $P^a$ and $m$ denote the axial current, pseudo-scalar density and degenerate quark mass, respectively. Here, we derive a Ward identity suggested in \cite{Bhattacharya:2005rb}, see \cite{Sint:1999ke} for an earlier account of similar identities.

We consider a singlet composite operator $\mathcal{O} = S^0 \mathcal{O}_\mathrm{ext}$ where the flavour singlet scalar density is defined by $S^0(x)=\mathrm{i}\bar\psi(x)T^0\psi(x)\,, T^0=-\mathrm{i}/\sqrt{2N_\mathrm{f}}\mathds{1}_{\Nf\times \Nf}$ in the $SU(N_\mathrm{f})$ flavour basis. The diagonal flavour matrix $T^0$ is introduced to handle flavour singlet quantities such as this one. With $\mathcal{O}_\mathrm{ext}$ lying outside $R$, we have $\langle \delta_A \mathcal{O}\rangle = \langle [\delta_A S^0(y)]\mathcal{O}_\mathrm{ext}\rangle$ and obtain
\begin{align}
\scalemath{0.85}{\delta_A S^0(y)= \frac{\mathrm{i}}{\sqrt{2\Nf}} \left[\delta_A \bar{\psi}(y) \psi(y) + \bar{\psi}(y) \delta_A\psi(y)\right]
= - \sqrt{\frac{2}{\Nf}}\epsilon^a(y) P^a(y)}
\end{align}
for the current under small axial variations $\epsilon^a(x)$ of the action. Inserting our choice of operator in eq.~(\ref{general_axial}) and following the same notation and procedure as in \cite{Heitger:2021bmg} (for a non-singlet Ward identity to obtain the parameter $Z$, defined below eq.~(\ref{gs:WI})), we transfer the Ward identity to the lattice:
\begin{subequations}
\label{eq:ward_long}
\begin{align}
&\scalemath{0.9}{Z_\mathrm{A} Z_\mathrm{S}^0 a^6 \sum_\mathbf{x,y} \langle \left[A_0^a(t_2;\mathbf{x}) - A_0^a(t_1;\mathbf{x})\right] S^0(y)\mathcal{O}^a\rangle}\\[-4pt]
&\scalemath{0.9}{- 2amZ_\mathrm{A} Z_\mathrm{S}^0 a^6 \sum_\mathbf{x,y} \sum_{x_0=t_1}^{t_2} \omega(x_0) \langle P^a(x)S^0(y)\mathcal{O}^a\rangle}\\[-4pt]
&\scalemath{0.9}{= - \sqrt{\frac{2}{\Nf}}Z_\mathrm{P} a^3 \sum_\mathbf{y}\langle P^a(y)\mathcal{O}^a\rangle}
\end{align}
\end{subequations}
where the external source was set to $\mathcal{O}_\mathrm{ext}=\mathcal{O}^a$ as defined in \cite{Heitger:2021bmg} and $Z_\mathrm{A}$, $Z_\mathrm{S}^0$ and $Z_\mathrm{P}$ are the renormalisation parameters. The weight factor $\omega(x_0)$ is needed to implement the trapezoidal rule for discretising integrals. Repeated flavour indices $a$ are summed over and $m$ is the PCAC mass.

We also consider another choice of operator for $\mathcal{O}$ in eq.~(\ref{general_axial}), $\mathcal{O}=\mathds{1}\mathcal{O}_\mathrm{ext}=\mathds{1}\mathcal{O}^a$. The Ward identity can be derived in the same fashion leading to
\begin{subequations}
\label{eq:ward_short}
\begin{align}
&\scalemath{0.9}{Z_\mathrm{A} Z_\mathrm{S}^0 a^6 \sum_\mathbf{x,y} \langle \left[A_0^a(t_2;\mathbf{x}) - A_0^a(t_1;\mathbf{x})\right] \mathcal{O}^a\rangle\langle S^0(y)\rangle}\\[-4pt]
&\scalemath{0.9}{- 2am Z_\mathrm{A} Z_\mathrm{S}^0 a^6 \sum_\mathbf{x,y}\sum_{x_0=t_1}^{t_2} \omega(x_0) \langle P^a(x)\mathcal{O}^a\rangle\langle S^0(y)\rangle}\\[-4pt]
&\scalemath{0.9}{= 0}
\end{align}
\end{subequations}
The main difference is that the r.h.s.~of eq.~(\ref{general_axial}) is zero in this case as $\langle \delta_A \mathcal{O}\rangle =   \langle [\delta_A \mathds{1}]\mathcal{O}_\mathrm{ext} \rangle + \langle \mathds{1}[\delta_A\mathcal{O}_\mathrm{ext}]\rangle=0 $: the flavour variation of the constant unity matrix is clearly zero (while the variation is generally zero outside $R$ where $\mathcal{O}_\mathrm{ext}$ is again defined).
 
Now it will become clearer why these two Ward identities (\ref{eq:ward_long}) and (\ref{eq:ward_short}) are of use. Having inserted the improvement pattern for all operators, a divergence cancels when the two equations are subtracted. The (cubic) divergence lies in the $e_\mathrm{S}(g_0^2)$ term in the improvement pattern of the flavour singlet scalar density \cite{Sint:1999ke} (see \cite{Bhattacharya:2005rb} for a notation more similar to the one applied here),
\begin{align}
(S_\mathrm{I})^0(x)=S^0(x)+\frac{1}{\sqrt{2\Nf}}a^{-3}e_\mathrm{S}(g_0^2)\mathds{1}+\frac{1}{\sqrt{2\Nf}}a\gs(g_0^2)\widetilde{\Tr}[F_{\mu\nu}(x)F_{\mu\nu}(x)]  +  \mathrm{O}(a^2)\label{S0_impr}\,,
\end{align}
where $\widetilde{\Tr}$ stands for the trace over colour indices.
Note that the discretised form of the field strength, $\widetilde{\Tr}[F_{\mu\nu}(x)F_{\mu\nu}(x)]$, must be local, symmetric w.r.t.~the site where the bilinear $S^0$ is placed and based on the gauge action employed, see \cite{Bhattacharya:2005rb}. We write this discretised form as $\{\widetilde{\Tr}[F F](x)\}^{ S_\mathrm{g}}$.

We apply the improvement scheme to $S^0$ in both Ward identities (\ref{eq:ward_long}) \& (\ref{eq:ward_short})  and subtract the now $\mathrm{O}(a)$ improved Ward identities from one another as to make use of the fact that $\langle a^{-3}e_\mathrm{S}(g_0^2)\mathds{1} \mathcal{O}\rangle =\langle a^{-3}e_\mathrm{S}(g_0^2)\mathds{1}\rangle \langle \mathcal{O}\rangle$. So all terms containing  $e_\mathrm{S}(g_0^2)$ are the same in both equations and cancel. The combined Ward identity simplifies to
\begin{subequations}
	\label{gs:WI}
	\begin{align}
	a^6 \sum_{x,\mathbf{y}}(\delta_{x_0,t_2}-\delta_{x_0,t_1})\bigg\{&\Big[\big\langle A_0^a(x) S^0(y)\mathcal{O}^a\big\rangle-\big\langle A_0^a(x) \mathcal{O}^a\big\rangle\big\langle S^0(y)\big\rangle\Big]\\
	+\frac{ag_\mathrm{S}}{\sqrt{2\Nf}}&\Big[\big\langle A_0^a(x) \{\widetilde{\Tr}[F F](y)\}^{ S_\mathrm{g}}\mathcal{O}^a\big\rangle-\big\langle A_0^a(x) \mathcal{O}^a\big\rangle\big\langle\{\widetilde{\Tr}[F F](y)\}^{ S_\mathrm{g}} \big\rangle\Big]\\
	+ac_\mathrm{A}\partial{x_0}&\Big[\big\langle P^a(x) S^0(y)\mathcal{O}^a\big\rangle-\big\langle P^a(x) \mathcal{O}^a\big\rangle\big\langle S^0(y)\big\rangle\Big]\bigg\}\\
	-a^6\sum_{\mathbf{x},\mathbf{y}} 2am \sum_{x_0=t_1}^{t_2}\omega(x_0)\bigg\{&\Big[\big\langle P^a(x)S^0(y)\mathcal{O}^a \big\rangle-\big\langle P^a(x)\mathcal{O}^a \big\rangle\big\langle S^0(y)\big\rangle\Big]\\
	+\frac{ag_\mathrm{S}}{\sqrt{2\Nf}}&\Big[\big\langle P^a(x)\{\widetilde{\Tr}[F F](y)\}^{ S_\mathrm{g}}\mathcal{O}^a \big\rangle-\big\langle P^a(x)\mathcal{O}^a \big\rangle\big\langle \{\widetilde{\Tr}[F F](y)\}^{ S_\mathrm{g}}\big\rangle\Big]\bigg\}\\
	=- &\sqrt{\frac{2}{\Nf}}a^3 \sum_\mathbf{y}Zr_\mathrm{m}\langle P^a(y)\mathcal{O}^a\rangle+\mathrm{O}(am,a^2)\,,
	\end{align}
\end{subequations}
having divided by $Z_AZ_S^0$ and employed $r_\mathrm{m}=\frac{Z_\mathrm{S}}{Z_S^0}$ and $Z=\frac{Z_\mathrm{P}}{Z_\mathrm{A}Z_\mathrm{S}}$.
Hence, the only unknown quantity left is $g_\mathrm{S}$; we already determined $Z$ and $r_\mathrm{m}$ in \cite{Heitger:2021bmg} and $c_\mathrm{A}$ in \cite{Bulava:2015bxa}. Note that quark-line disconnected diagrams appear in some of the three-point functions as a consequence of the flavour singlet property of the Ward identity considered here. Our determinations of $Z$ and $r_\mathrm{m}$ are based on flavour non-singlet Ward identities; thus no such disconnected diagrams entered there.

\section{Numerical Setup}
\label{sect:numerical-setup}
\begin{table}[h]
	\centering
		\begin{tabular}{lllccrl}
		\toprule
		$L/a$ &$\beta$ & $\kappa$ & \#REP &$N_\mathrm{sep}$ [MDU]  & $N_\mathrm{cnfg}$  & $\phantom{-}am^\mathrm{impr}$\\ 
		\midrule
		12 &3.4014 & $0.1368240$ & 4 &8 & 10466 & $\phantom{-}0.00065(16)$\\
		\midrule
		16 & 3.5522 &$0.1371379$ & 4 &8 & 11992 &$\phantom{-}0.000056(67)$ \\
		\midrule
		20 & 3.6900 &  $0.1371452$ & 10 &4 & 10000 &$\phantom{-}0.000094(59)$  \\ 
		\midrule
		24 &3.8013 & $0.1370387$ & 2 &8 & 8000  &$-0.000008(35)$ \\ 
		\midrule
		32 &3.9764 & $0.1367450$ & 2 &4 & 10072 & $\phantom{-}0.000011(14)$\\
		\bottomrule 
	\end{tabular}
	\caption{Simulation parameters $L$($=T$), $\beta$, $\kappa$, the number of replica \#REP and  $N_\mathrm{sep}$, the number of molecular dynamics units, MDU, all configurations are separated by. The ensembles are labelled by their lattice extent i.e. L12, L16, L20, L24, L32, and were originally generated as part of another study \cite{Fritzsch:2018yag,Conigli:2023rod}; we have (approximately) doubled the statistics of the ensembles L12 and L16. For each ensemble we list the $\mathrm{O}(a)$ improved PCAC mass $am^\mathrm{impr}$ obtained by averaging over the local masses of the three central time slices.} 
	\label{gs:tab:gauge_parameters_am}
\end{table}
We employ the tree-level Symanzik-improved gauge action with $N_\mathrm{f}=3$ mass-degenerate $\mathrm{O}(a)$ improved Wilson fermions. For the corresponding improvement coefficient $c_\mathrm{sw}$ we use the non-perturbative determination of \cite{Bulava:2013cta}. As for our earlier determination of $Z$ and $r_\mathrm{m}$ \cite{Heitger:2021bmg} we impose Schr\"odinger functional boundary conditions at the temporal boundaries of the lattice since this setup is well-suited for massless renormalisation schemes.
All gauge field ensembles used in this study are summarised in table~\ref{gs:tab:gauge_parameters_am} and lie on a precisely tuned line of constant physics (LCP), defined by a fixed spatial and temporal extent of $L=T\approx1\,$fm and (nearly) massless quarks, see table~\ref{gs:tab:gauge_parameters_am}. For the tuning, the gradient flow running coupling was kept constant, see \cite{DallaBrida:2016kgh,Conigli:2023rod,Fritzsch:2018yag}; the ensembles were originally generated as part of another study \cite{Fritzsch:2018yag, Conigli:2023rod} and we have (approximately) doubled the statistics of the ensembles L12 and L16.  The LCP ensures that our estimates of $\gs$ become smooth functions 
of the lattice spacing, higher-order ambiguities vanishing monotonically.  For all ensembles we use one-loop boundary $\mathrm{O}(a)$ improvement for both the gauge and fermion fields (i.e. the appropriate $\ct, \cttil$ values).  For full $\mathrm{O}(a)$ improvement of the Schrödinger functional correlation functions we additionally require the improvement coefficient $c_\mathrm{A}$, non-perturbatively known from \cite{Bulava:2015bxa}. Consequently, the quantity of interest, $\gs$, is the only unknown quantity in our numerical calculations (based on eq.~(\ref{gs:eq:gS_WI_explicit_ij})) as we make use of our results for $Z$ and $r_\mathrm{m}$ from \cite{Heitger:2021bmg}.

Most of the needed correlation functions needed overlap with those already implemented and evaluated in \cite{Heitger:2021bmg}. Due to the now flavour singlet nature of the Ward identities, two additional types of correlation functions are relevant: we implement single propagator traces via stochastic estimators. To reduce the variance at moderate cost we combine frequency splitting, to treat intermediate modes, and a hopping parameter expansion, to treat the high modes, as proposed in \cite{Giusti:2019kff}. We incorporate the necessary gluonic contributions in an {\ttfamily{openQCD}} based code \cite{openqcd}, as anticipated below eq.~(\ref{S0_impr}), including all contributing rectangles and plaquettes symmetrically w.r.t.~the site of the bilinear.

The Markov chain Monte Carlo sampling of the gauge field configurations suffers from critical slowing down of the topological charge for smaller lattice spacings. As in \cite{Heitger:2021bmg}, we project the data to the trivial topological sector (as suggested in \cite{Fritzsch:2013yxa}) so that the insufficient sampling of all sectors is no longer relevant: Ward identities hold in one sector only as well, irrespective of the sector at hand.
The statistical error analysis is carried out with  \texttt{pyerrors}, a \texttt{python} implementation \cite{Joswig:2022qfe} of the $\Gamma$-method \cite{Wolff:2003sm}, combined with linear error propagation via automatic differentiation \cite{Ramos:2018vgu,Schaefer:2010hu}. 

\section{Analysis}

We write down the $g_\mathrm{S}$ Ward identity introduced in eq.~(\ref{gs:WI}) in terms of the Schrödinger functional correlation functions with explicit flavour indices $i,j$ not summed over; note that the Wick contractions are not written out here. We obtain\footnote{Note that we no longer use the conventions from \cite{Bhattacharya:2005rb} but those employed in \cite{DallaBrida:2023fpl} because of the way correlation functions are defined in our {\ttfamily{openQCD}}  based codes. In the former a hermitian $F_{\mu\nu}$ is used while an anti-hermitian definition is utilised in the latter case. We still label the improvement parameter of the flavour singlet density by $\gs$ from  \cite{Bhattacharya:2005rb}, as before, by replacing $\ds$, used in the latter convention \cite{DallaBrida:2023fpl}, by $-gs$ as $\gs=-\ds$.}
\begin{subequations}
	\label{gs:eq:gS_WI_explicit_ij}
	{\allowdisplaybreaks
	\begin{align}
	\scalemath{0.92}{
	\def\Nf{\mathbf{N_{\rm f}}
	} \def\tr{\mathrm{tr}}}
	&\scalemath{0.92}{\left[f_\mathrm{AS}^{ij,\mathrm{con}}(t_2,y_0) -\Nf f_\mathrm{AS}^{ij,\mathrm{disc}}(t_2,y_0)\right]
	-\left[f_\mathrm{AS}^{ij,\mathrm{con}}(t_1,y_0) -\Nf f_\mathrm{AS}^{ij,\mathrm{disc}}(t_1,y_0)\right]}\label{gs:eq:gS_WI_explicit_ija}\\
	\scalemath{0.92}{+}&\scalemath{0.92}{\left[f_\mathrm{A}^{ij}(t_2)-f_\mathrm{A}^{ij}(t_1) \right]\Nf f_{\langle \mathrm{S}\rangle}^{ii}(y_0)}\\
	\scalemath{0.92}{-}&\scalemath{0.92}{g_\mathrm{S} \left\{2\left[f^{ij,\mathrm{disc}}_\mathrm{AE}(t_2,y_0)-f^{ij,\mathrm{disc}}_\mathrm{AE}(t_1,y_0)\right]
		-2[f_\mathrm{A}^{ij}(t_2)-f_\mathrm{A}^{ij}(t_1)]f_{\langle \mathrm{E}\rangle}(y_0)\right\}}\\
	\scalemath{0.92}{+}&\scalemath{0.92}{c_\mathrm{A} \bigg \{ a\partial_{t_2}	 \left[ f_\mathrm{PS}^{ij,\mathrm{con}}(t_2,y_0) -\Nf f_\mathrm{PS}^{ij,\mathrm{disc}}(t_2,y_0)\right]
	-a\partial_{t_1}		\left[f_\mathrm{PS}^{ij,\mathrm{con}}(t_1,y_0) - \Nf f_\mathrm{PS}^{ij,\mathrm{disc}}(t_1,y_0)\right]}\\
	&\scalemath{0.92}{\phantom{++} + \left[ a\partial_{t_2}f_\mathrm{P}^{ij}(t_2)-a\partial_{t_1}f_\mathrm{P}^{ij}(t_1)\right]\Nf f_{\langle \mathrm{S}\rangle}^{ii}(y_0) \bigg \}}\\
	-&\scalemath{0.92}{2am\left\{\left[\tilde{f}_\mathrm{PS}^{ij,\mathrm{con}}(t_1,t_2,y_0) -\Nf \tilde{f}_\mathrm{PS}^{ij,\mathrm{disc}}(t_1,t_2,y_0)\right]
	+\tilde{f}_\mathrm{P}^{ij}(t_1,t_2) \Nf f_{\langle \mathrm{S}\rangle}^{ii}(y_0)\right.}\label{gs:eq:gS_WI_explicit_ijf}\\
	&\scalemath{0.92}{\phantom{+++}-\left.g_\mathrm{S}\,\cdot\,2\left[\tilde{f}^{ij,\mathrm{disc}}_\mathrm{PE}(t_1,t_2,y_0) -\tilde{f}_\mathrm{P}^{ij}(t_1,t_2) f_{\langle \mathrm{E}\rangle}(y_0)\right]\right\}}\\
	\scalemath{0.92}{=}&\scalemath{0.92}{-2 Zr_\mathrm{m}f_\mathrm{P}^{ij}(y_0)+\mathrm{O}(am,a^2)}\,,
	\end{align}}
\end{subequations}
\hspace{-0.16cm}where `disc' refers to quark-line disconnected contributions. In this notation the explicit $\Nf$ dependence, a consequence of the flavour singlet nature of the identity, becomes apparent: Clearly, only the quark-line disconnected diagrams are multiplied by $\Nf$, in particular those that involve $S^0$ and not those that are purely gluonic ($f_{\langle \mathrm{E}\rangle}$). Thus, regarding the Wick contractions this can be understood as an additional quark loop for each additional quark flavour. Except for the quark-line and gluonic disconnected contributions, this flavour singlet identity is not that different from the flavour non-singlet case considered in \cite{Heitger:2021bmg} for the determination of $Z$: The connected correlation functions are the same, i.e. with the same Wick contractions, disregarding different prefactors which result from the different flavour structures. Thus the definitions of the above correlation functions can be found in the appendices of \cite{Heitger:2021bmg}. Note that `con' superscripts were not used then because of the lack of disconnected contributions. We list the additional correlation functions needed here: 
\begin{align}
\scalemath{0.9}{f_\mathrm{AS}^{ij,\mathrm{disc}}(x_0,y_0)
=-\frac{1}{2}\dfrac{a ^{12}}{L^3} \sum_{\mathbf{x,y, u, v}} \Bigg \langle \Big \langle \bar \psi_i(x) \gamma_0 \gamma_5 \psi_j(x)  \,\cdot \, \bar \zeta_j({\mathbf u}) \gamma_5 \zeta_i({\mathbf v})\Big \rangle_\mathrm{F} \cdot \Big \langle \bar \psi_j(y)\mathds{1}\psi_j(y) \Big \rangle_\mathrm{F} \Bigg \rangle_G}\,.
\end{align}
Substituting $\gamma_0\gamma_5$ by $\gamma_5$ we define $f_\mathrm{PS}^{ij,\mathrm{con}}(x_0,y_0)$ and $f_\mathrm{PS}^{ij,\mathrm{disc}}(x_0,y_0)$ analogously. For the same analysis, we also need to establish
\begin{align}
\scalemath{0.9}{f^{ij,\mathrm{disc}}_\mathrm{AE}(x_0,y_0)
=-\frac{1}{2}\frac{a^{9}}{L^3} \sum_{\mathbf{x,y, u, v}} \Bigg \langle  \Big \langle \bar \psi_i(x) \gamma_0\gamma_5 \psi_j(x) \, \cdot \, \bar \zeta_j({\mathbf u})\gamma_5\zeta_i({\mathbf v})\Big \rangle_\mathrm{F} \cdot \Big \langle E(y) \Big\rangle_\mathrm{F}\Bigg \rangle_G}\,.
\end{align}
and the purely gluonic case of $f_{\langle \mathrm{E}\rangle}(y_0) = a^4/2\sum_{\mathbf y}\langle\widetilde{\Tr}[FF](y) \rangle_G$. The correlation functions $\tilde{f}_\mathrm{PS}^{ij,\mathrm{con}}(t_1,t_2,y_0)$, $\tilde{f}_\mathrm{PS}^{ij,\mathrm{disc}}(t_1,t_2,y_0)$ and $\tilde{f}_\mathrm{PE}^{ij,\mathrm{disc}}(t_1,t_2,y_0)$ are defined by
\begin{align}
\scalemath{0.9}{
\tilde{f}^{ij}(t_1,t_2,y_0)=\sum_{x_0=t_1}^{t_2} \omega(x_0) f^{ij}(x_0,y_0) \quad \mathrm{where} \quad  \omega(x_0) =
\begin{cases}
\frac{1}{2},& \text{for } x_0=t_1,t_2\\
1,              & \text{otherwise}
\end{cases}}
\label{eq:summed-f-y0}
\end{align}
setting, e.g. $f^{ij}(x_0,y_0) = f_\mathrm{PS}^{ij,\mathrm{con}}(t_1,t_2,y_0)$ for $\tilde{f}_\mathrm{PS}^{ij,\mathrm{con}}(t_1,t_2,y_0)$. The single propagator trace contributing is defined by $
f_{\langle \mathrm{S}\rangle}^{ii}(y_0)=a^3\sum_{\mathbf y} \big \langle \bar \psi_i(y)\mathds{1} \psi_i(y) \big\rangle\,.$
\begin{figure}
	\centering
	\includegraphics[width=0.45\linewidth]{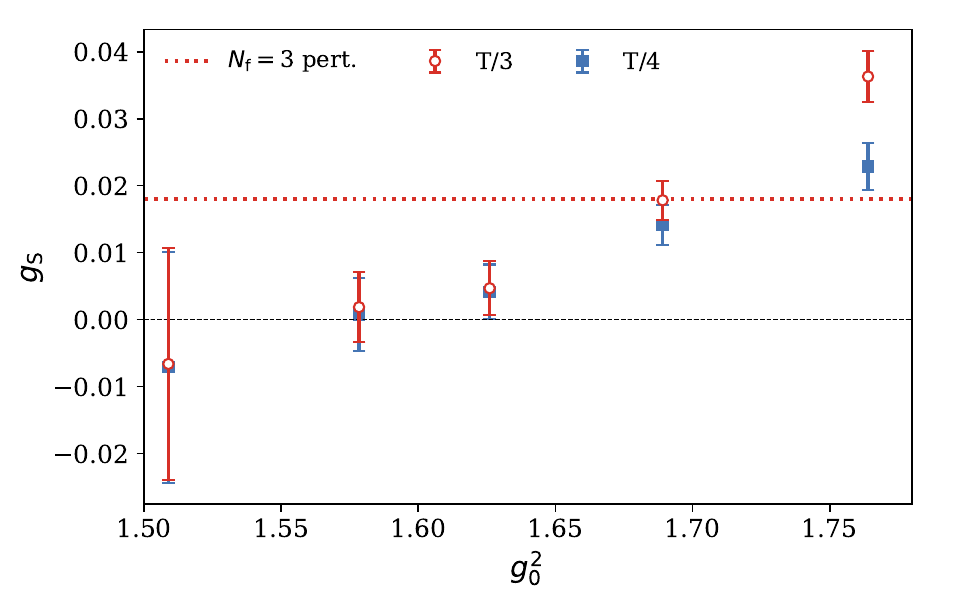}	
	\caption{$g_\mathrm{S}$ results obtained via eq.~(\ref{gs:eq:gS_WI_explicit_ij}) for $\Nf=3$ using our $Z$ and $r_\mathrm{m}$ interpolation formulas from \cite{Heitger:2021bmg} to construct the $Zr_\mathrm{m}$ values at the given couplings needed. The perturbative prediction from eq.~(\ref{eq:gs-pertubative}) (dashed red line) cannot be directly compared to the $\gs$ data points, see text below. The time interval $[t_1,t_2]$ (in eq.~(\ref{gs:eq:gS_WI_explicit_ij})) is set to $[T/4,3T/4]$ (labelled by T/4) or $[T/3,2T/3]$ (labelled by T/3) as shown by different colours.}
	\label{gs:fig:gS_vs_g02}
\end{figure}

We evaluate the Ward identity for $\Nf=3$ on the ensembles summed up in table~\ref{gs:tab:gauge_parameters_am}: by working close to the chiral limit with (nearly) vanishing quark masses (see table~\ref{gs:tab:gauge_parameters_am}), $\rmO(am)$ effects may be safely neglected and the Ward identity becomes valid up to $\rmO(a^2)$ cut-off effects. In \cite{Heitger:2021bmg} we found that for our $Z$ Ward identity the term above proportional to the current quark mass $m$, which can in principle be discarded in the chiral limit, still stabilises the chiral extrapolation (also found in refs.~\cite{Bulava:2016ktf,Heitger:2020mkp,Heitger:2020zaq,Chimirri:2023ovl} previously). Thus we prefer to keep it here too (even though we do not perform chiral extrapolations here but extract the Ward identity in the (nearly) chiral limit). We evaluate the operators $P^a$ and $S^0$ at  $t_1 \approx T/3$ and $t_2 \approx 2T/3$ (labelled by $T/3$) respectively, or alternatively at $t_1 \approx T/4$ and $t_2 \approx 3T/4$  (labelled by $T/4$). In practice, we round $t_1$ and $t_2$ to the nearest integer when $T/3$ and $T/4$ are not integers. We made the same choices previously for the determination of $Z$ in \cite{Heitger:2021bmg}. We extract the necessary PCAC mass plateaus by averaging over the local mass results of the three central time slices, see table~\ref{gs:tab:gauge_parameters_am} for the results at each coupling.

Putting everything together and solving the Ward identity for $\gs$, we obtain $\gs$ estimates at five lattice spacings and for two choices of the time interval labelled $T/3$ and $T/4$, shown in fig.~\ref{gs:fig:gS_vs_g02}. We find that both $\gs$ estimates approach zero for small couplings and are even compatible with zero within their respective $1\sigma$-errors at the two smallest couplings available; $\gs$ at the next higher coupling is close to zero as well. We also display the prediction from perturbation theory. Note that this perturbative prediction needs to be treated with caution in the present context since it is derived by applying $\bg =2g_0^2\gs$ to the perturbative prediction of $\bg$, cf. eq.~(\ref{eq:gs-pertubative}). As we will discuss in the next section the $\gs$ estimates here cannot be used in this relation to obtain $\bg$: for eq.~(\ref{bg-gs-relation}) to hold our non-perturbative definition of $\gs$ through eq.~(\ref{gs:eq:gS_WI_explicit_ij}) needs an additional term, which is merely an $\mathrm{O}(a)$ ambiguity for $\gs$. Therefore, the perturbative prediction can only give us an idea of the order of magnitude of $\gs$; as can be inferred from fig.~\ref{gs:fig:gS_vs_g02} our $\gs$ estimates differ from the perturbative prediction but the order of magnitude agrees well.



\section{Conclusion and Outlook}
\label{sect:conclusion-and-outlook}

We were able to determine non-perturbative estimates for $\gs$ for the first time. They were determined employing the tree-level Symanzik-improved gauge action with $N_\mathrm{f}=3$ mass-degenerate $\mathrm{O}(a)$ improved Wilson fermions and lie in the range of the CLS couplings, $g_0^2\in[1.5,1.77]$.

As to be able to make use of the $\gs$--$\bg$ relation and extract $\bg$ estimates from our $\gs$ estimates additional steps are necessary. As derived in appendix A of \cite{DallaBrida:2023fpl} via gradient flow observables the derivative of the full lattice action w.r.t.~the coupling $g_0^2$ enters the derivation of the $\gs$--$\bg$ relation. In the action used here not only the gauge action depends on the coupling but also the $\mathrm{O}(a)$ improved fermionic part of the action, through the improvement coefficient $\csw(g_0^2)$. This introduces an additional term $\propto \csw(g_0^2)$ in the derivation which was possibly overlooked in the original publication \cite{Bhattacharya:2005rb}. For the $\gs$--$\bg$ relation to be preserved in the $\mathrm{O}(a)$ improved theory, this term must also be taken into account in the choice of discretisation used for $\Tr (F_{\mu\nu} F_{\mu\nu})(x)$ in the improvement of the flavour singlet density, introduced in eq.~(\ref{S0_impr}). This amounts to  \cite{DallaBrida:2023fpl}
\begin{align}
\scalemath{0.92}{\widetilde{\Tr}[F_{\mu\nu}(x)F_{\mu\nu}(x)]  \rightarrow \bigg(\{\widetilde{\Tr}[F F](x)\}^{ S_\mathrm{g}}+ ag_0^2 \times \frac{\partial \csw({g_0^2})}{\partial g_0^2} \frac{\mathrm{i}}{2}   \mathcal{O}^\mathrm{clover}(x)\bigg)}
\label{clover-term}
\end{align}
where $\{\widetilde{\Tr}[F F](x)\}^{ S_\mathrm{g}}$ is the discretisation only based on the gauge action, used to determine $\gs$ here. The additional term is the derivative w.r.t.~$g_0^2$ of the fermionic part of the full $\mathrm{O}(a)$ improved action used here, introducing a dependence on the clover term.

Therefore, so that the $\gs$ estimates can be related to $\bg$ via $\bg =2g_0^2\gs$ the additional, clover term like expression needs to be included too, i.e.~the Ward identity from eq.~(\ref{gs:WI}) needs to be reevaluated with $\{\widetilde{\Tr}[F F](x)\}^{ S_\mathrm{g}}$ replaced by eq.~(\ref{clover-term}). More precisely, we add to the l.h.s.~of eq.~(\ref{gs:eq:gS_WI_explicit_ij}) two terms of the form (where `C' stands for the insertion of the clover term like expression)
\begin{align}
&\scalemath{0.92}{
-\gs\bigg\{\left[f_\mathrm{AC}^{ij,\mathrm{con}(t_2,y_0)}-N_\mathrm{f}f_\mathrm{AC}^{ij,\mathrm{disc}}(t_2,y_0)\right]-\left[f_\mathrm{AC}^{ij,\mathrm{con}}(t_1,y_0)-N_\mathrm{f}f_\mathrm{AC}^{ij,\mathrm{disc}}(t_1,y_0)\right]\bigg\}},\\
&\scalemath{0.92}{+2am\cdot \gs\bigg\{ \left[\tilde{f}_\mathrm{PC}^{ij,\mathrm{con}}(t_1,t_2,y_0)-N_\mathrm{f}\tilde{f}_\mathrm{PC}^{ij,\mathrm{disc}}(t_1,t_2,y_0)\right]+\tilde{f}_\mathrm{P}^{ij}(t_1,t_2)N_\mathrm{f}f_{\langle \mathrm{C} \rangle}^{ii}(y_0) \bigg\}}
\end{align}
similar to eq.~(\ref{gs:eq:gS_WI_explicit_ija}) and  eq.~(\ref{gs:eq:gS_WI_explicit_ij}f), respectively, i.e.~with the clover term like part from eq.~(\ref{clover-term}) instead of $S^0$ and with an additional factor, namely $\gs$.
Despite its gluonic component, $f_{\langle \mathrm{C} \rangle}(y_0)$ can be evaluated similarly to $f_{\langle \mathrm{S}\rangle}^{ii}(y_0)$ (cf. sect.~\ref{sect:numerical-setup}) since the expectation value can be rearranged such that the same single propagator trace $S(x,x)$  appears,
\begin{equation}
\scalemath{0.92}{
f_{\langle \mathrm{C} \rangle }(x_0)\sim-\frac{a^3}{2} \sum_{\mathbf{x}}\big\langle S(x,x)_{\beta\alpha}^{ba}\left(\sigma_{\mu\nu}F_{\mu\nu}^\mathrm{clover}\right)^{ab}_{\alpha\beta}(x)\big\rangle = -\frac{a^3}{2} \sum_{\mathbf{x}}\big\langle \eta^k, \left(\sigma_{\mu\nu}F_{\mu\nu}^\mathrm{clover}\right)^{kl} D^{-1}\eta^l\big\rangle}
\end{equation}
where $\eta^k(\eta^l)$ is the $k(l)$th (colour, Dirac) component of a stochastic noise source, $k(l)=1,...,12$ and $a,b$ denote colour and $\alpha,\beta$ Dirac indices. The term $\left(\sigma_{\mu\nu}F_{\mu\nu}^\mathrm{clover}\right)$ is a $12\times12$ matrix ($3\times3$ colour matrix times $4\times4$ Dirac matrix). 
In practice this clover Dirac-colour matrix is computed on every lattice point and multiplied by the stochastic estimator,  which the inverse Dirac operator, $D^{-1}$, is applied to before. This implementation is the most challenging part and is currently being tested; the coding of the other additional diagrams is more straight forward. Once all these additional diagrams are implemented, we plan to redetermine $\gs$ from the altered Ward identity.
The new $\gs$ estimates are expected to differ from the former in cut-off effects. We then plan to apply $\bg =2g_0^2\gs$ to extract non-perturbative $\bg$ estimates in the range of the CLS couplings, $g_0^2\in[1.5,1.77]$. 
Ultimately, we then intend to fit $\bg(g_0^2)$ as a smooth function of $g_0^2$ including the one-loop perturbative constraint from eq.~(\ref{eq:bg-pertubative}) as to arrive at an interpolation formula. We plan to repeat the same procedure for the altered $\gs$ using the perturbative constraint from eq.~(\ref{eq:gs-pertubative}).

\acknowledgments
We thank Anastassios Vladikas for participating in this project in its early stages. We also thank Rainer Sommer and Stefan Sint for useful discussions. This work was supported by the U.S. Department of Energy contract DE-AC05-06OR23177, under which Jefferson Science Associates, LLC operates Jefferson Lab (P. L. J. P.) and by the Deutsche Forschungsgemeinschaft (DFG) through the Research Training Group ``GRK 2149: Strong and Weak Interactions -- from Hadrons to Dark Matter'' (P. L. J. P., J. H., F. J.). Computations for this publication were performed on the HPC cluster PALMA II of the University of Münster.
\bibliographystyle{JHEP}
\bibliography{bib}

\end{document}